\documentclass[aps,twocolumn,showkeys,showpacs]{revtex4}
\usepackage{epsf}

\begin{document}

\title{Flux Tube Model Signals for Baryon Correlations\\
       in Heavy Ion Collisions}
\author{Apoorva Patel}
\email{adpatel@cts.iisc.ernet.in}
\affiliation{CHEP and SERC, Indian Institute of Science,
             Bangalore-560012, India}
\date{\today}

\begin{abstract}
The flux tube model offers a pictorial description of what happens during
the deconfinement phase transition in QCD. The three-point vertices of a
flux tube network lead to formation of baryons upon hadronization. Therefore,
correlations in the baryon number distribution at the last scattering
surface are related to the preceding pattern of the flux tube vertices in
the quark-gluon plasma, and provide a signature of the nearby deconfinement
phase transition. I discuss the nature of the expected signal, and how to
extract it from the experimental data for heavy ion collisions at RHIC and
LHC.
\end{abstract}

\maketitle

\input{axodraw.sty}
\section{Introduction}

The theory of strong interactions, QCD, incorporates the non-perturbative
phenomena of confinement and chiral symmetry breaking. These phenomena
are expected to disappear at high temperatures and/or at large chemical
potential, where QCD can be analysed reliably because its effective
coupling becomes weak. How the phenomena arise in low temperature and low
chemical potential region is crucial to our understanding of the QCD
dynamics, ranging from hadronization in the early universe to the interior
of neutron stars. A lot of effort has been devoted towards this goal,
experimentally through the study of heavy ion collisions, theoretically
through construction of phenomenological models (since we have not been
able to solve QCD accurately), and computationally through simulations
of lattice QCD. The results are often summarized as the phase structure
of QCD in the temperature-chemical potential space. Recent reviews of the
subject are available in
Refs.\cite{alford,kharzeev,mclerran,loizides,sgupta,kanaya}.

\begin{figure}
\begin{center}{
\setlength{\unitlength}{0.8mm}
\begin{picture}(100,80)
  \thicklines
  \put(30,10){\vector(1,0){70}}
  \put(30,10){\vector(0,1){60}}
  \put(30,10){\vector(-1,2){20}}
  \put(30,50){\line(1,0){60}}
  \put(15,40){\line(1,0){60}}
  \put(15,80){\line(1,0){60}}
  \put(15,40){\line(0,1){40}}
  \put(75,40){\line(0,1){40}}
  \put(90,10){\line(0,1){40}}
  \put(30,50){\line(-1,2){15}}
  \put(90,50){\line(-1,2){15}}
  \put(90,10){\line(-1,2){15}}

  \put(15,65){\circle*{2}}
  \put(30,30){\circle*{2}}
  \put(60,10){\circle*{2}}
  \put(30,10){\circle*{1}}
  \put(15,40){\circle*{1}}

  \put(15,65){\line(1,0){60}}
  \qbezier{(15,65),(18,60),(20,52)}
  \qbezier{(75,65),(77,60),(78,53)}
  \qbezier[60]{(20,52),(50,52),(78,53)}
  \multiput(20,62)(5,0){12}{\circle{1}}
  \multiput(21,58)(5,0){12}{\circle{1}}
  \multiput(22,54)(5,0){12}{\circle{1}}

  \put(30,30){\line(-1,2){2.5}}
  \qbezier[40]{(27.5,35),(40,38),(75,40)}
  \qbezier{(30,30),(60,30),(60,10)}
  \qbezier{(60,10),(57,28),(75,40)}
  \multiput(57,37)(5,0){3}{\makebox(0,0)[bl]{$\ast$}}
  \multiput(34,34)(5,0){7}{\makebox(0,0)[bl]{$\ast$}}
  \multiput(31,31)(5,0){7}{\makebox(0,0)[bl]{$\ast$}}
  \multiput(48,28)(5,0){3}{\makebox(0,0)[bl]{$\ast$}}
  \multiput(55,25)(5,0){2}{\makebox(0,0)[bl]{$\ast$}}
  \put(57,22){\makebox(0,0)[bl]{$\ast$}}
  \put(58,18.5){\makebox(0,0)[bl]{$\ast$}}

  \put(95,5){\makebox(0,0)[bl]{$\mu$}}
  \put(32,70){\makebox(0,0)[bl]{$T$}}
  \put(0,40){\makebox(0,0)[bl]{$m=\infty$}}
  \put(15,5){\makebox(0,0)[bl]{$m=T=\mu=0$}}
  \put(5,20){\makebox(0,0)[bl]{$m_{\rm phys}\Longrightarrow$}}
\end{picture}
}\end{center}
\vspace{-5mm}
\caption{Schematic description of the phase structure of QCD in the
$m-T-\mu$ space. First-order transition surfaces are shown shaded, and
critical lines are shown dotted. Value of $m$ corresponding to the real
world QCD is indicated by an arrow. Color superconductor phases occurring
at large chemical potential are omitted.}
\label{fig:phasestruct}
\end{figure}
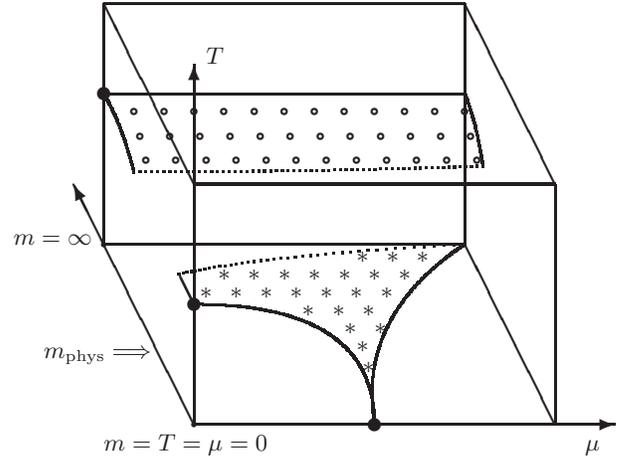

Consider QCD with $N$ colors and $N_f$ degenerate quark flavors of mass $m$,
at temperature $T$ and quark chemical potential $\mu$. The phase structure of
this theory is depicted in Fig.\ref{fig:phasestruct} in a schematic manner.
(More details are needed to describe the real world QCD. Complexities arising
from unequal quark masses and color superconductor phases at large chemical
potential are omitted here.) Certain boundaries of the phase structure are
better understood than the interior region, because of the exact symmetries
present there. Explicitly:\\
(1) For $m=\infty$, the pure gauge theory has a finite temperature
deconfinement phase transition, governed by the breaking of the global
$Z_N$ center symmetry of the Polyakov loop. This transition is of first
order for $N\geq3$.\\
(2) For $m=0=\mu$, the theory has a finite temperature chiral phase
transition, governed by the restoration of the flavor $SU(N_f)_V$
symmetry to $SU(N_f)_L \otimes SU(N_f)_R$. This transition is of first
order for $N_f\geq3$.\\
(3) For $m=0=T$, the theory has a baryon condensation phase transition,
where the vacuum structure changes from $\langle\bar\psi\psi\rangle\neq0$
to $\langle\psi^\dagger\psi\rangle\neq0$. This transition occurs roughly
when $\mu$ equals the constituent quark mass, and is also of first order.

First-order phase transitions are generically stable against small
changes of symmetry breaking perturbations. So the above three phase
transitions extend inwards, to varying extent, from the boundaries of
the phase structure. The behavior expected from phenomenological models
and numerical calculations is that the first-order transition surfaces
end in critical lines, as shown in Fig.\ref{fig:phasestruct}.
Specifically, for the physical values of the quark masses, there is no
phase transition as $T$ is varied, unless $\mu$ is sufficiently large.
In the analytic cross-over region connecting high and low temperature
regions, the three nearby transitions mentioned above cause various QCD
properties to change rapidly. In general, which property is affected how
much by which transition depends on the property concerned. To verify all
the theoretical expectations, therefore, it is important to construct
appropriate experimental observables that highlight the individual aspects
of deconfinement, chiral symmetry restoration and baryon condensation,
when QCD is tested at high temperature and/or at large chemical potential.


In this article, I focus on the signals in heavy ion collision experiments
related to the deconfinement phase transition. The collisions produce a high
energy non-equilibrium state of QCD matter, which subsequently expands and
cools.  The experimental signatures observed at RHIC and LHC demonstrate
that the initial fireball first (quasi-)equilibrates to a quark-gluon
plasma (QGP), which then hadronizes and decays. The detected quantities
are multiplicities and distributions of various types of hadrons. Since a
sizeable fraction of the colliding nuclei continues unscattered along the
beam direction, the signals coming from the fireball are cleanly observable
only in sufficiently transverse directions, say for pseudo-rapidity $|\eta|
< \eta_{\rm max}$. These signals can be broadly separated into two types:
(a) those describing the initial non-equilibrium fireball state and its
approach to equilibrium, and (b) those characterizing the properties of an
equilibrated QGP. The former consist of particles that can manage to pass
through the QGP, e.g. direct photons and leptons, heavy quark jets, high
$p_T$ jets and the elliptic flow. The latter consist of distributions of
moderate $p_T$ hadrons produced close to the surface of the fireball, and
they are the ones I want to relate to properties of the QGP.

The QGP is a strongly interacting medium, while the hadrons resulting from
its decay do not interact much with each other. The hadronization process
thus goes through a stage where the scattering events among the components
emerging from the fireball drop rapidly. Afterwards, the hadrons essentially
propagate radially outward without scattering, although the unstable ones
decay. From the observed detector signals, it is possible to backtrack the
distributions of hadrons to this ``kinetic freeze-out" stage of the fireball.
This stage is analogous to the ``last scattering surface" in the evolution of
the cosmic microwave background radiation (CMBR), and several methodologies
developed to study the fluctuations there are useful in its analysis.

It should be kept in mind that the QGP is strongly coupled in the cross-over
region, and the decoupling of hadrons goes through several stages instead of
being instantaneous \cite{heinz}. Hadronization of the (quasi-)equilibrated
QGP takes place at energy density $\epsilon_{\rm cr}\simeq1 {\rm GeV/fm}^3$
and $T_{\rm cr}\simeq175$MeV. Shortly thereafter, inelastic scattering of
hadrons stops, resulting in ``chemical freeze-out" when $T_{\rm chem}\simeq
170$MeV. Upon further expansion, elastic and resonant scattering (mediated
largely by pions) ceases, producing ``kinetic freeze-out" when $T_{\rm kin}
\simeq120$MeV. The scatterings contribute significantly to the thermalization
of the hadron momenta, but assuming that the hadronic medium has low diffusion,
we can see through them to the patterns in the QGP.

The distribution of hadrons at the kinetic freeze-out stage has often been
modeled as a thermalized hadron resonance gas. Such thermal models fit the
observed multiplicities of various hadrons, but miss all the multi-particle
correlations between hadrons. The strongly interacting QGP should have many
correlations, and more accurate models are therefore needed to relate them
to observable signals. In what follows, I describe how a flux tube model
providing a physical picture of the deconfinement process predicts specific
two-particle correlations in the distribution of hadrons, going beyond the
single particle multiplicities. I first present a summary of the model,
and then discuss the baryon number correlations predicted by it.

\section{The Flux Tube Model \cite{patel1}}

\subsection{Phenomenology}

The flux tube model of QCD is motivated by the dual superconductor
description of linear color confinement \cite{ripka}, where condensation
of color magnetic charges restricts color-electric fields to vortex-like
configurations. Although an exact derivation of the model from QCD has
not been found, the model describes strong coupling expansions in lattice
QCD and has been phenomenologically quite successful. A characteristic
property of the flux tube is its energy per unit length, i.e. the string
tension $\sigma$. Other than that, the flux tube has a finite width $w$ and
a persistence length $a$ (arising from the tube's stiffness so that the flux
tube has to go a certain distance before it can freely reorient itself),
both assumed to be of order $\Lambda_{QCD}^{-1}$.

The flux tubes have to obey the constraint of Gauss's law. So they terminate
only on quarks, and interact only at $N$-point vertices. These two features
represent the invariant tensors $\delta_{ab}$ and $\epsilon_{abc}$ (in case
of $SU(3)$) used to describe the meson and the baryon wavefunctions. Other
multi-quark hadron states are phenomenologically not prominent, except for
multi-nucleon nuclei, and so all other interactions among the flux tubes
are ignored in the model \cite{foot1}. Note that a glueball would be
represented by a closed flux tube loop in this description.

\begin{figure}
\begin{center}{
\setlength{\unitlength}{1pt}
\begin{picture}(250,160)
  \thicklines
  \put(50,120){\makebox(0,0)[bl]{(a)}}
  \Photon(10,150)(100,150){5}{1}
  \put(55,151){\vector(-2,1){4}}
  \put(10,150){\circle*{4}} \put(100,150){\circle*{4}}
  \put(10,130){\makebox(0,0)[bl]{$\overline{Q}$}}
  \put(100,130){\makebox(0,0)[bl]{$Q$}}

  \put(180,120){\makebox(0,0)[bl]{(c)}}
  \Photon(135,150)(195,150){5}{1}
  \Photon(165,150)(225,150){5}{1}
  \put(150,156){\vector(-1,0){2}} \put(210,146){\vector(-1,0){2}}
  \put(180,156){\vector( 1,0){2}} \put(180,146){\vector( 1,0){2}}
  \put(135,150){\circle*{4}} \put(225,150){\circle*{4}}
  \put(135,130){\makebox(0,0)[bl]{$\overline{Q}$}}
  \put(225,130){\makebox(0,0)[bl]{$Q$}}

  \put(50,-10){\makebox(0,0)[bl]{(b)}}
  \put(75,100){\circle{20}}
  \put(77,90.5){\vector(4,1){4}}
  \Photon(10,50)(40,50){30}{1}
  \Photon(40,50)(80,50){40}{1}
  \Photon(80,50)(100,50){30}{0.5}
  \put(40,50){\vector(-1,-4){2}}
  \put(10,50){\circle*{4}} \put(100,50){\circle*{4}}
  \put(10,30){\makebox(0,0)[bl]{$\overline{Q}$}}
  \put(100,30){\makebox(0,0)[bl]{$Q$}}

  \put(180,-10){\makebox(0,0)[bl]{(d)}}
  \Photon(145,0)(145,100){3}{1.5} \Photon(170,0)(175,105){-3}{2}
  \Photon(200,10)(195,95){3}{1}   \Photon(225,0)(225,100){2}{2}
  \Photon(125,80)(143,90){1}{0.5} \Photon(125,20)(143,10){-1}{0.5}
  \Photon(142,79)(173,81){2}{0.5} \Photon(148,50)(170,44){-2}{0.5}
  \Photon(142,19)(155,20){1}{0.5}
  \Photon(172,95)(199,75){-1}{1}  \Photon(174,16)(196,25){2}{1}
  \Photon(196,90)(226,95){2}{1}   \Photon(199,68)(224,71){-1}{1}
  \Photon(196,40)(226,30){2}{0.5} \Photon(215,71)(215,55){-1}{0.5}
  \Photon(224,55)(240,50){2}{0.5} \Photon(224,20)(240,30){-2}{0.5}
  \put(150,21){\vector(4,1){2}}   \put(145,31){\vector(1,4){2}}
  \put(159,46){\vector(-4,1){2}}  \put(170,35){\vector(1,-4){2}}
  \put(186,21){\vector(-1,0){2}}  \put(196.5,33){\vector(0,1){2}}
  \put(199,54){\vector(-1,-4){2}} \put(209,69){\vector(4,1){2}}
  \put(215,65){\vector(0,1){2}}
  \put(155,20){\circle*{4}} \put(215,55){\circle*{4}}
  \put(155,25){\makebox(0,0)[bl]{$\overline{Q}$}}
  \put(215,45){\makebox(0,0)[bl]{$Q$}}
\end{picture}
}\end{center}
\caption{Possible flux tube configurations connecting a static quark-antiquark
pair, as the temperature is increased (from top to bottom), and when baryonic
vertices are included (from left to right).}
\label{fig:fluxtubes}
\end{figure}
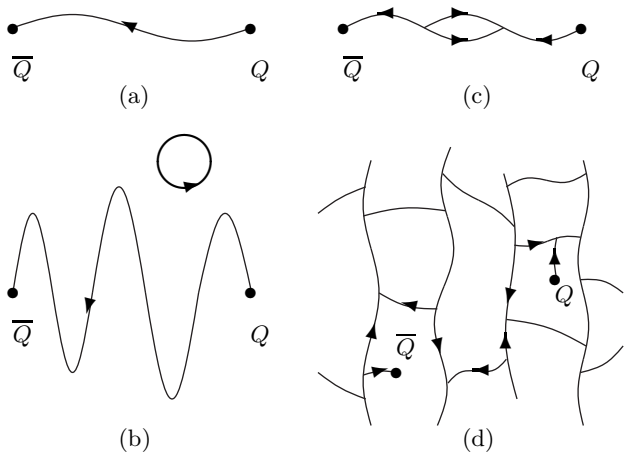

The finite temperature behavior of the model is governed by the competition
between the energy and the entropy of the flux tube configurations. At low
temperatures, energy wins, keeping down the total length of the flux tubes.
At high temperatures, entropy dominates, producing elaborate structures of
the flux tubes all over the space.

First consider the situation for the pure gauge theory with only static
color sources, i.e. $m=\infty$. As the temperature is increased the flux
tubes oscillate more, and also produce more vertices. Some of the possible
configurations are shown in Fig.\ref{fig:fluxtubes}. In absence of vertices,
corresponding to the $SU(2)$ gauge theory, there is a second-order
deconfinement phase transition when the flux tube length diverges and
the quark-antiquark pair loses information about each-other's position.
In presence of vertices, corresponding to $SU(N>2)$ gauge theories, the flux
tubes can percolate the space in a network before their length can diverge.
That also allows the quark-antiquark pair (hooked on to the network) to
lose information about each-other's position, but produces a first-order
deconfinement phase transition.

\begin{figure}
\begin{center}{
\setlength{\unitlength}{0.8mm}
\begin{picture}(100,50)
  \thicklines
  \put( 0,45){\makebox(0,0)[bl]{(a)}}
  \put(15,45){\line(1,0){75}}
  \put(55,45){\vector(-1,0){5}}
  \put(15,45){\circle*{2}} \put(90,45){\circle*{2}}
  \put(10,45){\makebox(0,0)[bl]{$\overline{Q}$}}
  \put(95,45){\makebox(0,0)[bl]{$Q$}}

  \put( 0,30){\makebox(0,0)[bl]{(b)}}
  \put(15,30){\line(1,0){35}}  \put(55,30){\line(1,0){35}}
  \put(35,30){\vector(-1,0){5}} \put(75,30){\vector(-1,0){5}}
  \put(15,30){\circle*{2}} \put(90,30){\circle*{2}}
  \put(50,30){\circle*{1}} \put(55,30){\circle*{1}}
  \put(10,30){\makebox(0,0)[bl]{$\overline{Q}$}}
  \put(95,30){\makebox(0,0)[bl]{$Q$}}
  \put(50,32){\makebox(0,0)[bl]{$q$}}
  \put(55,32){\makebox(0,0)[bl]{$\overline{q}$}}

  \put( 0,10){\makebox(0,0)[bl]{(c)}}
  \put(15,10){\line(1,0){35}}  \put(60,10){\line(1,0){30}}
  \put(60,10){\line(-1,2){5}}  \put(60,10){\line(-1,-2){5}}
  \put(35,10){\vector(-1,0){5}} \put(75,10){\vector(-1,0){5}}
  \put(55,20){\vector(1,-2){3}} \put(55,0){\vector(1,2){3}}
  \put(15,10){\circle*{2}} \put(90,10){\circle*{2}}
  \put(55,0){\circle*{1}} \put(50,10){\circle*{1}} \put(55,20){\circle*{1}}
  \put(10,10){\makebox(0,0)[bl]{$\overline{Q}$}}
  \put(95,10){\makebox(0,0)[bl]{$Q$}}
  \put(52,10){\makebox(0,0)[bl]{$q$}}
  \put(52,20){\makebox(0,0)[bl]{$q$}}
  \put(52,0){\makebox(0,0)[bl]{$q$}}
\end{picture}
}\end{center}
\caption{A color-electric flux tube can break when dynamical quarks
are included in the theory. (a) A flux tube produced by static color
sources. (b) Its breaking by a quark-antiquark pair appearing from the
vacuum. (c) Its breaking by a baryon appearing from the vacuum at finite
chemical potential.}
\label{fig:fluxtubebreaking}
\end{figure}
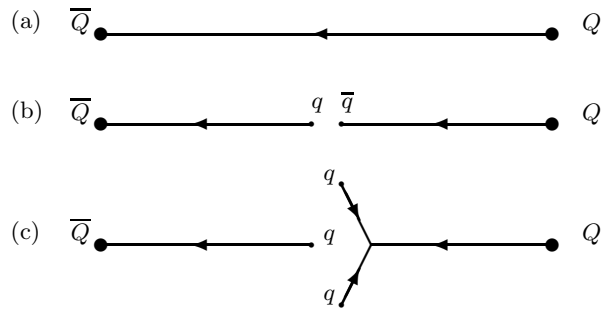

When finite mass quarks are included in the model, they can break the flux
tubes by quark-antiquark pair production from the vacuum. Baryon number
production at finite chemical potential can also break the flux tubes.
These possibilities are illustrated in Fig.\ref{fig:fluxtubebreaking}.
With the flux tube network breaking up, the strength of the deconfinement
phase transition weakens as the quark mass is lowered from $m=\infty$
and/or as the chemical potential is increased. Numerical estimates show
that for $N_f=3$ QCD at small chemical potentials, the first-order
deconfinement phase transition ends in a critical line around $m=1.5$GeV,
as depicted in Fig.\ref{fig:phasestruct}. Although the cross-over region
does not have any sharp behavior of the deconfinement phase transition,
we can still investigate whether the percolating flux tube scenario
suggests any detectable signal there.

\subsection{Quantitative Formulation}

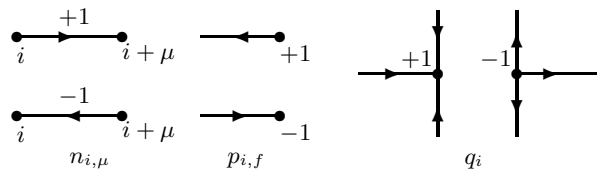
\begin{figure}[!b]
\begin{center}{
\setlength{\unitlength}{0.7mm}
\begin{picture}(110,30)
  \thicklines
  \put(0,10){\line(1,0){20}}        \put(0,25){\line(1,0){20}}
  \put(0,10){\circle*{2}}           \put(0,25){\circle*{2}}
  \put(20,10){\circle*{2}}          \put(20,25){\circle*{2}}
  \put(20,10){\vector(-1,0){11}}    \put(0,25){\vector(1,0){11}}
  \put(0,5){\makebox(0,0)[bl]{$i$}} \put(0,20){\makebox(0,0)[bl]{$i$}}
  \put(20,5){\makebox(0,0)[bl]{$i+\mu$}} \put(20,20){\makebox(0,0)[bl]{$i+\mu$}}
  \put(8,12){\makebox(0,0)[bl]{$-1$}} \put(8,27){\makebox(0,0)[bl]{$+1$}}

  \put(35,10){\line(1,0){15}}       \put(35,25){\line(1,0){15}}
  \put(50,10){\circle*{2}}          \put(50,25){\circle*{2}}
  \put(50,25){\vector(-1,0){9}}     \put(35,10){\vector(1,0){9}}
  \put(50,5){\makebox(0,0)[bl]{$-1$}} \put(50,20){\makebox(0,0)[bl]{$+1$}}

  \put(65,18){\line(1,0){15}}       \put(95,18){\line(1,0){15}}
  \put(80,18){\circle*{2}}          \put(95,18){\circle*{2}}
  \put(65,18){\vector(1,0){8}}      \put(95,18){\vector(1,0){8}}
  \put(73,19){\makebox(0,0)[bl]{$+1$}} \put(88,19){\makebox(0,0)[bl]{$-1$}}
  \put(80,6){\line(0,1){24}}        \put(95,6){\line(0,1){24}}
  \put(80,6){\vector(0,1){6}}       \put(95,18){\vector(0,-1){8}}
  \put(80,30){\vector(0,-1){6}}     \put(95,18){\vector(0,1){8}}

  \put(10,0){\makebox(0,0)[bl]{$n_{i,\mu}$}}
  \put(40,0){\makebox(0,0)[bl]{$p_{i,f}$}}
  \put(85,0){\makebox(0,0)[bl]{$q_i$}}
\end{picture}
}\end{center}
\caption{The link and site variables for the flux tube model.}
\label{fig:modelvariables}
\end{figure}

It is straightforward to formulate the model on a lattice with spacing $a$.
The flux tubes live on the links, while the quarks and vertices live on
the sites. The variables take values $0,\pm1$, depending on the direction
of the flux (for $N>2$), as shown in Fig.\ref{fig:modelvariables}. The
total energy of a flux tube configuration is
\begin{equation}
E = \sigma a \sum_{i,\mu} |n_{i,\mu}| + m \sum_{i,f} |p_{i,f}|
  + v \sum_i |q_i| ~,
\label{totalenergy}
\end{equation}
where $v$ denotes the energy cost of an $N$-point vertex, and $f$ sums over
the $2N_f$ spin and flavor quark degrees of freedom. For a cubic lattice,
the constraint of Gauss's law at every site becomes
\begin{equation}
\sum_\mu (n_{i,\mu} - n_{i-\mu,\mu}) - \sum_f p_{i,f} + N q_i
\equiv \alpha_i = 0 ~.
\label{gausslaw}
\end{equation}
When Gauss's law is applied globally, the contribution of the flux variables
$n_{i,\mu}$ cancels. We then have two ways of determining the total baryon
number $B$, either from the quark variables $p_{i,f}$ or from the vertex
variables $q_i$:
\begin{eqnarray}
&& \sum_i \alpha_i = N \sum_i q_i - \sum_{i,f} p_{i,f} = 0 \nonumber \\
&& \Longrightarrow B = {1 \over N} \sum_{i,f} p_{i,f} = \sum_i q_i ~.
\label{baryonno}
\end{eqnarray}

The grand canonical partition function for the system is
\begin{equation}
Z[T,\mu] = \sum_{n_{i,\mu},~p_{i,f},~q_i} \exp\left[
         - {E - \mu NB \over T} \right] ~ \prod_i \delta_{\alpha_i,0} ~,
\label{partfn1}
\end{equation}
with Eq.(\ref{baryonno}) implying that quark chemical potential $\mu$ is
equivalent to vertex chemical potential $N\mu$. The grand canonical partition
function is fully factorized by expressing the Kronecker delta constraint at
every site as
\begin{equation}
\delta_{\alpha_i,0} = \int_{-\pi}^\pi {d\theta_i \over 2\pi}
                    ~ e^{i\alpha_i\theta_i} ~.
\label{deltafn}
\end{equation}
The sum over the variables $n_{i,\mu},~p_{i,f},~q_i$ can then be carried out
explicitly, resulting in
\begin{eqnarray}
Z[T,\mu] &=& \int_{-\pi}^\pi \prod_i {d\theta_i \over 2\pi} ~
             \prod_{i,\mu} (1 + 2e^{-\sigma a/T} \cos(\theta_{i+\mu}-\theta_i))
             \nonumber \\
    &\times& \prod_i \left(1 + 2e^{-m/T}
                     \cos\left(\theta_i + i{\mu\over T}\right)\right)^{2N_f}
             \nonumber \\
    &\times& \prod_i (1 + 2e^{-v/T} \cos(N\theta_i)) ~.
\label{partfn2}
\end{eqnarray}
The equivalence of quark and vertex chemical potentials, following from
Eq.(\ref{baryonno}), makes $Z[T,\mu]$ invariant under the global symmetry
transformation, $\theta_i\rightarrow\theta_i+\epsilon$, for any complex
value of $\epsilon$.

By extending the allowed values of the variables $n_{i,\mu},~p_{i,f},~q_i$
to all integers \cite{foot2}, the grand canonical partition function can be
converted to \cite{patel2}
\begin{eqnarray}
Z[T,\mu] &=& \int_{-\pi}^\pi \prod_i {d\theta_i \over 2\pi}
          ~  \exp\left[ J \sum_{i,\mu} \cos(\theta_{i+\mu}-\theta_i)
          \right. \nonumber \\
         &+& \left. h \sum_i \cos\left(\theta_i + i{\mu\over T}\right)
                  + p \sum_i \cos(N\theta_i) \right] .
\label{partfn3}
\end{eqnarray}
The systems represented by Eq.(\ref{partfn2}) and Eq.(\ref{partfn3}) are both
in the universality class of the XY spin model in the presence of an ordinary
magnetic field as well as a $Z(N)$ symmetric magnetic field. Their phase
structures are similar, and their couplings are related by
\begin{equation}
J = 2e^{-\sigma a/T} ~,~~ h = 4N_f~e^{-m/T} ~,~~ p = 2e^{-v/T} ~.
\label{effcoupl}
\end{equation}
In quantitative analysis, Eq.(\ref{partfn3}) has the convenience that its
algebraic structure is suitable for treatment using well-established methods
in statistical mechanics.

The physical meaning of the site variables $\theta_i$ can be uncovered
by looking at the free energy of a static color charge in the system.
Introduction of a quark at site $j$ modifies the Gauss's law constraint
there as $\delta_{\alpha_j,0} \rightarrow \delta_{\alpha_j,-1}$. That makes
the free energy of a static quark
\begin{equation}
\exp(-F_q/T) = \langle \exp(-i\theta_j) \rangle ~.
\label{quarkenergy}
\end{equation}
In the finite temperature gauge field theory language, this free energy is
given by the expectation value of the Polyakov loop $P_j$ at the site $j$.
So we arrive at the correspondence that $\theta_i$ represents the phase of
the Polyakov loop $P_i$, and the flux tube description of deconfinement is
dual to the familiar Polyakov loop description of deconfinement \cite{patel3}.
The important advantage of the flux tube description is that it provides a
position space visual representation of what happens as the temperature is
varied in QCD.

Another advantage of the flux tube formulation is that it can be numerically
simulated without any fermion sign problem at finite chemical potential
\cite{detar,chandrasekharan,deforcrand,gattringer}, since Eq.(\ref{partfn1})
involves only real positive weights while Eq.(\ref{partfn3}) necessitates
complex weights.

\subsection{Limitations}

The flux tube model described above does not incorporate the chiral nature
of the quark degrees of freedom. So it cannot properly address the features
in the QGP cross-over region that can be considered consequences of the nearby
chiral phase transition. Attempts have been made to combine Polyakov loop
and Nambu-Jona-Lasinio models \cite{ogilvie,fukushima,ratti}, to study chiral
properties of the QGP. They have treated the Polyakov loop only as a mean
field without its full dynamics, but have still provided reasonable fits to
static QGP properties extracted from lattice QCD simulations. They need to be
extended by including the kinetic term for the Polyakov loop dynamics, e.g.
$J\ne0$ in Eq.(\ref{partfn3}), to obtain a good estimate of the fluctuations.
One of the predictions of chiral symmetry restoration at high temperature is
an increased production of baryons relative to mesons, due to a decrease in
the constituent quark mass. The same enhancement is also predicted by the
increase in the number of flux tube vertices at high temperature. This
coincidence makes the prediction robust, and it has indeed been observed
\cite{STAR}, but an estimate of its magnitude requires accurate treatment
of both confinement and chiral dynamics. That is an exercise for the future.

Without the chiral properties of quarks, the flux tube model is also unable
to say much about the baryon condensation phase transition. On the other hand,
influence of a finite chemical potential on the deconfinement phase transition
can be inferred from the behavior of the grand canonical partition function.  
The particle$\leftrightarrow$antiparticle symmetry of QCD corresponds to
invariance of $Z[T,\mu]$ under simultaneous sign flips of the variables
$n_{i,\mu},~p_{i,f},~q_i$. The partition function is therefore an even
function of $\mu$, i.e. $Z[T,\mu] = Z[T,-\mu]$. Convexity of the exponential
function then implies
\begin{equation}
Z[T,\mu] \geq Z[T,\mu=0] ~,~~ F[T,\mu] \leq F[T,\mu=0] ~.
\label{convexfn}
\end{equation}
$F[T,\mu]$ is continuous across the coexistence surface, and $dF = -SdT -Nd\mu$
at constant volume. So for positive latent heat, it follows that introduction
of a chemical potential decreases the transition temperature.

Furthermore, as described earlier, a non-zero chemical potential weakens the
signal from the deconfinement phase transition. On the other hand, it enhances
the signal from the baryon condensation phase transition, observable in baryon
number susceptibilities. Baryon condensation needs to be treated as a site
percolation problem and not as bond percolation by flux tubes. That will not
be addressed here.

\section{Baryon Number Correlation Signals}

Let us consider the flux tube scenario of what happens in the heavy ion
collision experiments as the fireball of the QGP expands and cools. Even
though the cross-over region relevant to experiments does not possess a
single percolating flux tube network, it can still contain many finite
clusters of flux tubes. Because the flux is directed, an obvious feature
of every such cluster is that any neighbor of a vertex is an anti-vertex
and vice versa (see Fig.2(d)). During the evolution of the fireball, the
total baryon number is conserved \cite{foot3}, and so the vertices can only
be locally pair-produced or pair-annihilated. As the QGP hadronizes, the
flux tube clusters start breaking up. After the chemical freeze-out stage,
there is no more production or annihilation of vertices; every vertex ends
up in a baryon and every anti-vertex ends up in an antibaryon. In the absence
of subsequent large-scale diffusion, the radial propagation of (anti)baryons
preserves the geometric pattern of (anti)vertices present at the chemical
freeze-out stage. As a result, the angular positions of the (anti)baryons
seen in the detector can be backtracked to the angular positions of the
(anti)vertices on the surface of the fireball at the chemical freeze-out
stage. This pattern of vertices on the surface of the fireball can then be
analysed for correlations and fluctuations, using techniques similar to
those used to analyse the temperature fluctuations in the cosmic microwave
background radiation \cite{weinberg}.

The above description is a simple consequence of picking the right variables
to visualize confinement dynamics in QCD. Although quantitative estimates of
the (anti)vertex distributions would require numerical simulations, there
are certain qualitative features that can be gleaned with much less effort.
The fact that heavy ion collisions produce a sizeable number of antibaryons
\cite{STAR}, from an initial state that has none, means that (a) a good
number of baryonic and antibaryonic vertices are produced in the fireball,
and (b) fragmentation of flux tubes during hadronization is more likely
than their shrinking and coalescing that would annihilate vertices with
anti-vertices. Our aim is to look for a specific pattern in the distribution
of the produced (anti)vertices. Note that the flux tube scenario is unable
to say anything regarding the energy and the momenta of the hadrons that
emerge from the fireball.

\subsection{Experimental Data Parametrization}

There are several differences in the type of data gathered from CMBR and
from heavy ion collision experiments, which need to be kept in mind when
studying the two with similar techniques.\\
(1) The CMBR is a single high statistics event where homogeneity and isotropy
of the universe allow accurate determination of the distribution parameters,
while heavy ion collisions are multiple modest statistics events where
ensemble averaging improves the accuracy of the distribution parameters.\\
(2) The temperature data of CMBR are real numbers, while the baryon number
data of heavy ion collisions are integers that need to be binned and smeared.\\
(3) The average temperature and size of the ``last scattering surface" are
well-determined for CMBR, while the number of participating nucleons and the
fireball volume fluctuate considerably in case of heavy ion collisions and
are not accurately determined.\\
(4) The CMBR data cover the full $4\pi$ solid angle and hence can be easily
parametrized using the orthogonal basis of the spherical harmonics, while the
heavy ion collision data are restricted to sufficiently transverse directions
only and a suitable orthogonal basis for its parametrization has to be found.

Let us assume that the available data cover $\theta_m<\theta<\pi-\theta_m$,
and label the angular distribution by the unit vector $\hat{n}$. The heavy
ion collision experiments have an axial symmetry around the beam axis as well
as the reflection (or parity) symmetry $\theta\leftrightarrow\pi-\theta$.
These symmetries are sufficient to prevent the average baryon number flow in
any direction, and can be used to improve the statistical accuracy of the
correlations in the data. Fourier expansion helps in search for correlations
by orthogonal separation of scales, and the baryon number distribution can be
partially expanded as
\begin{equation}
b(\hat{n}) \equiv b(\theta,\phi) = {1\over\sqrt{2\pi}} \sum_{\sigma=\pm}
                  \sum_{m=-\infty}^{\infty} b_m^{\sigma}(\theta) ~ e^{im\phi} ~.
\label{baryondist}
\end{equation}
Here $b_m^{\pm}(\theta) = \pm b_m^{\pm}(\pi-\theta)$, reality of
$b(\hat{n})$ implies $b_m^{\sigma *}(\theta) = b_{-m}^{\sigma}(\theta)$,
and $\langle b(\hat{n}) \rangle = b_0^+(\theta)/\sqrt{2\pi}$.
In a sense, the positive and the negative parity terms correspond to all the
even and all the odd values of $l$, respectively, of the spherical harmonics
expansion. Since the hadrons emerging from the fireball in longitudinal
directions go undetected, the total and detected baryon numbers of a
collision event are (note that $\tanh\eta=\cos\theta$):
\begin{eqnarray}
B_{\rm tot} &=& \sqrt{2\pi} \int_{-1}^1 d(\cos\theta) ~ b_0^+(\theta) ~, \\
B_{\rm det} &=& \sqrt{2\pi} \int_{-\cos\theta_m}^{\cos\theta_m}
              d(\cos\theta) ~ b_0^+(\theta) ~.
\label{baryonnum}
\end{eqnarray}

The two-point baryon number correlation function is the ensemble average
$\langle b(\hat{n})b(\hat{n}') \rangle$. Because of the axial symmetry,
it depends only on the difference of the azimuth angles $\phi-\phi'$,
and because of the reflection symmetry the product of the parities
$\sigma\sigma'$ has to be one. Therefore,
\begin{equation}
\langle b_m^{\sigma}(\theta) b_{-m'}^{\sigma'}(\theta') \rangle =
\langle b_m^{\sigma}(\theta) b_{m'}^{\sigma' *}(\theta') \rangle =
\delta_{\sigma\sigma'}\delta_{mm'} C_m^{\sigma}(\theta,\theta') ~,
\label{coeffortho}
\end{equation}
and the two-point correlation function becomes
\begin{equation}
\langle b(\hat{n}) b(\hat{n}') \rangle = {1\over2\pi}
                   \sum_{\sigma=\pm} \sum_{m=-\infty}^{\infty}
                   C_m^{\sigma}(\theta,\theta') ~ e^{im(\phi-\phi')} ~.
\label{baryoncor}
\end{equation}
Inverting this relation, we obtain
\begin{eqnarray}
C_m^{\sigma}(\theta,\theta') &=& {1\over2\pi}
            \int_0^{2\pi} \!d\phi \int_0^{2\pi} \!d\phi' ~ e^{im(\phi'-\phi)}
            \langle b^{\sigma}(\hat{n}) b^{\sigma}(\hat{n}') \rangle ~,
                             \nonumber \\
b^{\pm}(\hat{n}) &=& {1\over2}
                     \left(b(\theta,\phi) \pm b(\pi-\theta,\phi)\right) ~.
\label{twoptcoeff}
\end{eqnarray}
The coefficient functions $C_m^{\pm}(\theta,\theta')$ are real and symmetric.
They contain all the information about the two-point correlation function.
The non-trivial correlations are given by the connected contributions,
\begin{equation}
\langle b(\hat{n}) b(\hat{n}') \rangle_c =
\langle b(\hat{n}) b(\hat{n}') \rangle -
        \langle b(\hat{n}) \rangle \langle b(\hat{n}') \rangle ~,
\label{connectcor}
\end{equation}
\begin{equation}
[C_m^{\sigma}(\theta,\theta')]_c = C_m^{\sigma}(\theta,\theta') -
             \delta_{m0} ~ \delta_{\sigma+} ~ b_0^+(\theta) ~ b_0^+(\theta') ~.
\label{connectcoeff}
\end{equation}
In practice, depending upon the resolution available in the data,
(a) the expansion can be truncated at a suitable value of $|m|$, and
(b) the $\theta$-dependence can be further subdivided into smaller bins.

Similar parametrizations of the data can be carried out for distributions of
other conserved quantities also, e.g. the electric charge or the strangeness.
As a matter of fact, experimentally observed correlations for different
conserved quantities can be compared with each other, to illustrate different
aspects of the QGP dynamics. The flux tube model predictions are the cleanest
in case of the baryon number, and illustrate the deconfinement mechanism.
Different models with appropriate features would be required to understand
the dynamics of other conserved quantities.

It should be noted that the theoretical estimates of the correlations are
obtained assuming that the volume of the QGP is sufficiently large. On the
other hand, the number of participating nucleons in the collisions, and
hence the volume of the fireball, depends on the centrality of the heavy ion
collisions. So to determine as well as to control the consequent systematic
effects, it is necessary to determine the correlation functions in different
centrality ranges separately.

\subsection{Theoretical Expectations}

Theoretically, the equilibrium correlations between flux tube vertices are
straightforward to calculate in the three-dimensional position space, which
can then be projected onto the observed surface of the fireball. An alternating
neighbor pattern of vertices and anti-vertices, expected in a percolating
flux tube network, would give rise to two-point baryon number correlations
similar to the two-point charge correlations in ionic liquids (with ions of
comparable size) such as $Cs^+ Cl^-$. The key common ingredient in the two
cases is the hard-core repulsion between the objects involved.

\begin{figure}
\begin{center}{
\setlength{\unitlength}{0.8mm}
\begin{picture}(100,100)
  \thicklines
  \put(20,30){\vector(1,0){70}}
  \put(20,0){\vector(0,1){90}}

  \qbezier{(25,30),(35,30),(35,50)}
  \qbezier{(35,50),(38,120),(40,70)}
  \qbezier{(40,70),(46,30),(51,50)}
  \qbezier{(51,50),(55,70),(59,50)}
  \qbezier{(59,50),(63,40),(67,50)}
  \qbezier{(67,50),(71,60),(75,50)}
  \qbezier{(75,50),(79,45),(83,50)}
  \qbezier{(83,50),(87,54),(91,50)}

  \thinlines
  \qbezier{(25,30),(35,30),(35,15)}
  \qbezier{(35,15),(38,-20),(49,30)}
  \qbezier{(49,30),(55,50),(63,30)}
  \qbezier{(63,30),(68,20),(75,30)}
  \qbezier{(75,30),(83,48),(91,44)}

  \put(85,25){\makebox(0,0)[bl]{$r$}}
  \put(15,30){\makebox(0,0)[bl]{$0$}}
  \put(15,50){\makebox(0,0)[bl]{$1$}}
  \multiput(20,50)(2,0){35}{\circle*{0.5}}
  \put(12,85){\makebox(0,0)[bl]{$g(r)$}}
  \put(45,60){\makebox(0,0)[bl]{$g_{|v|}$}}
  \put(48,15){\makebox(0,0)[bl]{$g_{v}$}}
\end{picture}
}\end{center}
\caption{Schematic representation of the pair distribution functions
$g_{|v|}(r)$ (thick line) and $g_v(r)$ (thin line). The former is
similar to that for objects with hard-core repulsion. The latter is for a
percolating flux tube network where vertices and anti-vertices alternate.}
\label{fig:pairdistfn}
\end{figure}
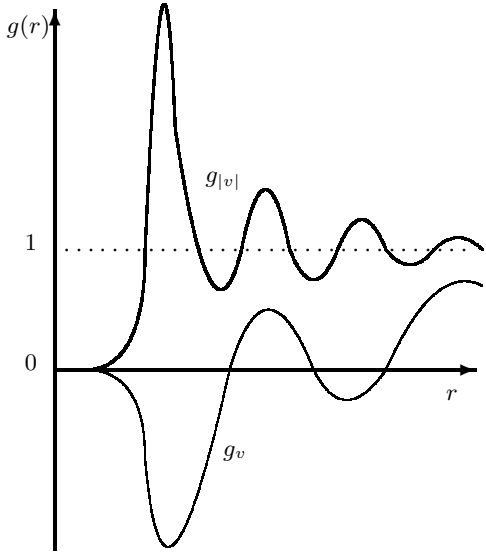

For discrete objects located at $\vec{r}_i$, a convenient description of the
position space correlations is in terms of the pair distribution function
\begin{equation}
\rho(\vec{r}) ~ g(\vec{r},\vec{r}') ~ \rho(\vec{r}') = \left\langle
                \sum_{i\ne j} \delta(\vec{r}-\vec{r}_i)
              ~ \delta(\vec{r}'-\vec{r}_j) \right\rangle ~,
\label{pairdist}
\end{equation}
where
$\rho(\vec{r})=\left\langle\sum_i\delta(\vec{r}-\vec{r}_i)\right\rangle$
is the local density of the objects. In homogeneous and isotropic fluids,
$\rho$ is independent of the position and $g$ depends only on
$|\vec{r}-\vec{r}'|$. Choosing $\vec{r}'=0$, then we have
\begin{equation}
\rho ~ g(r) = \left\langle \sum_{i\ne0}
              \delta(\vec{r}-\vec{r}_i) \right\rangle ~,
\label{paircorrel}
\end{equation}
Interactions fade away at long distances, and so asymptotically
$g(r\rightarrow\infty)=1$. For objects with no correlations, e.g. an ideal
gas, $g(r)=1$. For objects with hard-core repulsion, $g(0)=0$, and beyond the
hard core $g(r)$ tends to its asymptotic value exhibiting damped oscillations
\cite{hardcore}, as illustrated in Fig.\ref{fig:pairdistfn}. In particular,
the distance scale of the oscillations is determined by the inter-object
separation (the first peak corresponds to the likely nearest-neighbor
separation, the second peak corresponds to the likely next-nearest-neighbor
separation, and so on), and the amplitude of the oscillations is determined
by how tightly the objects are packed together.

To detect the correlations between vertices and anti-vertices, it is useful
to compare the pair distribution functions for $|q(\vec{r})|~|q(\vec{r}')|$
and $q(\vec{r})~q(\vec{r}')$, $g_{|v|}$ and $g_v$ respectively, on the same
data sets. Comparison of $g_{|v|}$ and $g_v$ keeps under control the effects
arising from variation in the number of participating nucleons in the heavy
ion collisions. Both follow the form of Eq.(\ref{pairdist}), with the sum
running over vertices and anti-vertices. The former omits any vertex value
signs and uses the density $\rho_{|v|}$, while the latter includes the vertex
values, i.e. $q_i q_j$ on the right-hand side of Eq.(\ref{pairdist}) and
$q_i$ in obtaining the density $\rho_v$ (note that $\rho_v < \rho_{|v|}$).

For a percolating flux tube network, the function $g_{|v|}(r)$ should behave
similar to that for hard-core objects, with the nearest-neighbor separation
of the order of the baryon size. Similar behavior would be expected from
several other fluid models also, so $g_{|v|}(r)$ can be treated as a
model-insensitive reference function. In contrast, in a percolating flux tube
network, successive neighbors contribute with opposite signs to $g_v(r)$.
The resultant pair distribution function would then have oscillations of the
form sketched in Fig.\ref{fig:pairdistfn}. For a system with little or no
correlations between vertices, e.g. single particle thermal distribution
models, the probability of occurrence of a vertex or an anti-vertex at any
location is proportional to its overall density. Contributions to the
right-hand side of Eq.(\ref{pairdist}) would then have the same density
factors as on the left-hand side, and $g_v(r)$ would behave the same way as
$g_{|v|}(r)$. Thus $g_v(r)$ is sensitive to the correlations present between
vertices and anti-vertices, and the contrast between $g_{|v|}(r)$ and $g_v(r)$
can be used as a measure of these correlations.

The percolating flux tube network in the high temperature QGP is maximally
connected for $m=\infty$, with $N$ nearest neighbors for each vertex and
anti-vertex. The network starts breaking up as the quark mass is lowered,
more so if the chemical potential is also increased. Even individual baryons
that are not part of any flux tube cluster can appear. Such a break up reduces
the correlations between neighbors, smoothening the oscillations of the pair
distribution function, analogous to what happens when a liquid is heated up.
For the physical values of the quark masses, there is no deconfinement phase
transition and no single percolating flux tube network is expected in the QGP.
Still, if large enough flux tube clusters survive in the cross-over region,
they would contribute to the contrast between $g_v(r)$ and $g_{|v|}(r)$.
The strength of the contrast, therefore, measures the extent to which the
nearby deconfinement phase transition influences the properties of the QGP
in the cross-over region.

In general, the first peak of the pair distribution function is the most
informative about the fluid properties. Its location provides an estimate
of the inter-object separation, and the area under it is a measure of the
number of nearest neighbors. In the present case, it is also the place
where the contrast between $g_v(r)$ and $g_{|v|}(r)$ is the maximum, and
so it is the best feature for identifying the correlations between vertices
and anti-vertices. Note that for specific values of the parameters in
Eq.(\ref{effcoupl}), numerical simulations of the flux tube model can
obtain the functions $g_v(r)$ and $g_{|v|}(r)$ for an equilibrated
homogeneous and isotropic QGP.

\subsection{Connecting Theory to Experiment}

Projection of the three-dimensional pair distribution function onto the surface
of the fireball smears the oscillatory structure of $g(r)$. The resultant
angular distribution can be expressed in terms of a smearing function as
\begin{equation}
w(\alpha) = \int_{r_{\rm min}}^{r_{\rm max}} S(\alpha,r) ~ g(r) ~ dr ~,~~
\int_0^{\pi} S(\alpha,r) ~ d\alpha = 1 ~.
\label{angsmear}
\end{equation}
Here the integral over $r$ has to be sufficiently restricted so as not to
lose the oscillatory signal. Since the correlations that we are interested
in are of short range, it is convenient to remove the uniform part of the
distribution function, e.g. look at $g(r)-1$ which describes the connected
part. In the context of picking up the vertex correlation signal, it is even
better to look at the difference $g_v(r)-g_{|v|}(r)$, which vanishes for
both $r\rightarrow0$ and $r\rightarrow\infty$.

In a specific setting, the smearing can be performed with sufficient accuracy
numerically. Approximate analytic result can be obtained, nevertheless, under
the assumption that the inter-vertex separation $r$ is much smaller than the
radius of the fireball $R$. Let $\vec{a}$ be the position of the center of
the pair with respect to the center of the fireball, and let $\beta$ be the
angle between $\vec{a}$ and $\vec{r}$. Then the pair separation $r$ projects
to angular separation $\alpha=(r/a)\sin\beta$. Taking expectation value over
$\beta$ and $a$, for a homogeneous and isotropic distribution of pairs in the
fireball, the smearing function becomes:
\begin{eqnarray}
\hspace{-5mm}
S(\alpha,r) &=& \int_0^R {3a^2 da \over R^3}
                \int_0^{\pi} {\sin\beta ~ d\beta \over 2} ~
                \delta\left(\alpha-{r \over a}\sin\beta\right) \\
            &=& \int_0^R {3a^2 da \over R^3} ~
                {\alpha a^2 \over r \sqrt{r^2-\alpha^2 a^2}} \nonumber\\
            &=& \cases{ 9\pi r^3/(16R^3 \alpha^4)
                        \quad: \alpha\geq r/R \cr
                        {3r^3 \over 32R^3 \alpha^4}
                        \left( 12\lambda - 8\sin(2\lambda)
                        + \sin(4\lambda) \right) \cr
                        \qquad\qquad\qquad\qquad\!:
                        \sin\lambda=\alpha R/r \leq 1}
\label{smearfn}
\end{eqnarray}
This small angle approximation to $S(\alpha,r)$ should be reasonable for
studying at least the first two peaks of $g_v(r)$, given that experimentally
the radius of the fireball in central heavy ion collisions is $\sim 6$fm and
the inter-nucleon separation in nuclear matter is $\sim 2$fm.

When the rotational symmetry is a good approximation, it is convenient to
decompose the angular distribution $w(\alpha)$ in terms of the orthogonal
Legendre polynomials:
\begin{eqnarray}
w(\alpha) &=& \sum_{l=0}^\infty C_l
              \left({2l+1 \over 4\pi}\right) P_l(\cos\alpha) ~,
              \nonumber \\
C_l &=& 2\pi \int_{-1}^1 d(\cos\alpha) ~ w(\alpha) ~ P_l(\cos\alpha) ~. 
\label{angdistfull}
\end{eqnarray}
An estimate of $w(\alpha)$ can then be compared to experimental data using
$\cos\alpha = \cos\theta\cos\theta' + \sin\theta\sin\theta'\cos(\phi-\phi')$,
but $C_l$ cannot be extracted from the experimental data unless full angular
coverage is available.
With only axial symmetry and limited angular coverage present, $w(\alpha)$
needs to be decomposed in terms of parity and azimuthal Fourier components
as in Eqs.(\ref{baryoncor},\ref{twoptcoeff}). Specifically, that avoids
systematic errors in the construction of the connected distribution
$w_c(\alpha)$. The addition theorem for the associated Legendre polynomials,
\begin{equation}
P_l(\cos\alpha) = \sum_{m=-l}^l (-1)^m
   P_l^{m}(\cos\theta) ~ P_l^{-m}(\cos\theta') ~ e^{im(\phi-\phi')} ~,
\label{Legendrereln}
\end{equation}
then relates the coefficient functions according to
\begin{eqnarray}
C_m^{\sigma}(\theta,\theta') &=& (-1)^m \\
    &\times& \mathop{{\sum}'}_{l=|m|}^{\infty} C_l \left({2l+1 \over 2}\right)
    P_l^m(\cos\theta) ~ P_l^{-m}(\cos\theta') ~, \nonumber
\label{coeffreln}
\end{eqnarray}
where the sum is restricted to even(odd) values of $l-|m|$ for $\sigma=+(-)$.
The functions $C_m^{\sigma}(\theta,\theta')$ can be extracted from theoretical
models as well as from the experimental data, and they can be restricted to
suitable ranges of $|m|$ and $\theta$ depending on the resolution available.

There still remain gaps between the theoretical formalism described here and
the actual experimental data. The most prominent one is the fact that the
detectors record only the charged hadrons. Thus protons and anti-protons are
observed but neutrons and anti-neutrons are not. The requisite baryon number
correlations can be extracted from the data only if the observed subset of
(anti)protons provides a faithful characterization of the total baryon number
distribution---ideally the two should be proportional. Moreover, corrections
need to be estimated due to only approximate equilibration of the fireball,
non-uniformity of the QGP caused by the elliptic flow, and baryon number
diffusion subsequent to hadronization, all of which are likely to weaken the
oscillatory correlation signal. On the other hand, development of the hard
baryon core during hadronization would enhance the oscillatory signal,
compared to the softer QGP state at higher temperature. Despite these gaps,
it is worthwhile to look for the two-point baryon number correlations in the
experimental data, as a characteristic signature of the deconfinement phase
transition. Specifically, experimental determination of the contrast between
$g_{|v|}$ and $g_v$, and their departure from $g(r)=1$ corresponding to no
correlations, would provide a model-independent characterization of the
baryon-antibaryon correlations.

\section{Summary and Outlook}

The Polyakov loop is a widely used order parameter for understanding the
finite temperature deconfinement phase transition in a gauge theory. 
Its dual description in terms of the flux tube model has the advantage
that the gauge theory dynamics can be visualized in position space.
This picture helps in connecting experimentally observable baryon number
correlations with properties of the deconfinement phase transition.
There is no fundamental interaction associated with the baryon number.
So the baryon number correlations have to arise from its precursors in
the QGP, i.e. the pattern of the flux tube vertices expected from QCD.

The experimental effort in studying heavy ion collisions has so far largely
focused on single particle distributions. In this article, I have gone beyond
that towards analysing multi-particle distributions. I have described how to
extract the two-point baryon number correlations from the experimental data,
and have also predicted that it would contain an oscillatory signal, based
on theoretically expected QGP properties at the chemical freeze-out stage.
The experimental data for moderate $p_T$ baryons in central heavy ion
collisions are the best suited for observation of the predicted signal.
The features to be quantified are the distance scale and the amplitude of
the oscillatory correlations. The former is expected to be the inter-baryon
separation ($\sim 2$fm), while the latter though uncertain at present would
tell us a lot about how tightly or softly the QGP is packed. It is worth
keeping in mind that the intense search for two-point correlations in the
CMBR, theoretically expected but without any accurate prediction of its
magnitude (inflation is a much more flexible theory than QCD), found the
correlations at the level of $\Delta T \approx 10^{-5}T$, and provided a
major boost to our understanding of cosmology.

As a different application, the flux tube picture may also help in improving
the baryon production estimates of hadronization models, by replacing the
creation of diquark pairs with the creation of baryonic vertex pairs.
It can also be noted that the flux tube dynamics influences diffusion in the
QGP. Specifically, formation of a flux tube network would suppress diffusion.
Both diffusion and viscosity characterize transport properties of a fluid,
incorporating dissipative dynamics and entropy production. The former is
convenient to use in position space, the latter in momentum space, and the
two can be related in a kinetic theory framework. Low diffusion is not
incompatible with the small viscosity to entropy density ratio in the high
entropy density QGP. Quantitative analysis of diffusion in the QGP requires
a hydrodynamic framework, however, which is beyond the scope of the formalism
discussed here.


\begin{thebibliography}{99}
\bibitem{foot1}{Many multi-quark hadron resonances are frequently described
      in terms of intermediate diquark components. The most attractive diquark
      configuration is the antisymmetric color-triplet one, which is covered
      by the $\epsilon_{abc}$ structure of the baryonic vertex.}
\bibitem{foot2}{The extension of the allowed values is fine for the bosonic
      variables $n_{i,\mu}$ and $v_i$, but it is not clear-cut for the
      fermionic variables $p_{i,f}$ due to the exclusion principle.}
\bibitem{foot3}{The experiments take place at fixed baryon number, while
      theoretical results are easier to obtain at fixed chemical potential.
      The two descriptions are related by a Legendre transformation.}
\end{thebibliography}
\end{document}